\documentclass[twocolumn,floatfix,nofootinbib,british,balancelastpage]{revtex4-1}

\usepackage{graphicx}
\usepackage{amsmath,amssymb,amsfonts}
\usepackage{multirow}
\usepackage{hyperref,enumitem}
\usepackage{bbm}
\usepackage{xspace}
\usepackage{verbatim}
\usepackage{bm}
\usepackage{color}
\usepackage{cancel}
\usepackage{mathtools}
\usepackage[utf8]{inputenc}

\usepackage[T1]{fontenc} 
\usepackage{slashed}
\usepackage{soul}
\usepackage{booktabs}
\usepackage{enumitem}
\usepackage{booktabs}
\usepackage{units}
\usepackage{subfigure}
\usepackage{fancyvrb}

\newcommand{\Checkmate}{\texttt{CheckMATE}}

\newcommand{\lint}{L_{\text{int}}}
\usepackage{soul}

\begin{document}
\fvset{samepage=true, fontsize=\scriptsize}


\title{Automatised ILC-Bounds on Dark Matter Models with CheckMATE \\ \vspace{0.75pt}\small{Talk presented at the International Workshop on Future Linear Colliders (LCWS2017), Strasbourg, France, 23-27 October 2017. C17-10-23.2.}}%

\author{Daniel Dercks}%
\email{daniel.dercks@desy.de}
\affiliation{II. Institut f\"ur Theoretische Physik, Universität Hamburg, Luruper Chaussee 149, 22761 Hamburg, Germany}

\author{Gudrid Moortgat-Pick}%
\email{gudrid.moortgat-pick@desy.de}
\affiliation{II. Institut f\"ur Theoretische Physik, Universität Hamburg, Luruper Chaussee 149, 22761 Hamburg, Germany}
\affiliation{DESY, Notkestra{\ss}e 85, D-22607 Hamburg, Germany}


\begin{abstract}
The public collider phenomenology computing tool CheckMATE (Check Models at Terascale Energies) was originally designed to allow theorists to quickly test their favourite BSM models against various existing LHC analyses performed by ATLAS and CMS. It offers an automatised chain of Monte Carlo event generation, detector simulation, event analysis and statistical evaluation so that it can automatically determine whether a given parameter point of a BSM model is excluded or not. Currently, it contains more than 50 individual ATLAS or CMS analyses whose several hundred signal regions target various final states as they typically appear in theories beyond the Standard Model. In this study, we extend this functionality to allow sensitivity studies for the International Linear Collider. As an example, we implement a dark matter monophoton search and use it to analyse three benchmark scenarios with different assumptions about the interaction between dark matter and Standard Model particles. We determine the ILC sensitivity expected for $\sqrt{s} = \unit[500]{GeV}, L_{\text{int}} = \unit[500]{fb}^{-1}$ and compare the results  for the cases of completely unpolarised beams and for individual lepton polarisation settings.
\end{abstract}


\date{\today}%
\maketitle

\section{Introduction}
The Large Hadron Collider (LHC) has been running very successfully in the last few years and the nearly \unit[60]{fb}$^{-1}$ of data taken at $\sqrt{s} = 8$ and $\unit[13]{TeV}$ has provided physicists around the world with an enormous amount of new information regarding the physics of proton-proton interactions. Complementarily, the International Linear Collider (ILC) will give deep insights into the underlying physics accessible with polarised electrons and positrons at high energies. 

At both LHC and ILC, a large fraction of these searches are specifically designed to find hints of a supersymmetric extension of the Standard Model (SM). However, this theory --- though well motivated as it answers many open questions of the SM --- is clearly not the only feasible theory beyond the Standard Model and there exists a very large number of possibilities to extend the particle spectrum and/or to formulate new interactions between particles. Unfortunately, it is practically impossible for a particle physics experiment to analyse the data and interpret the results in the context of each individual of those theories. Moreover, an interesting new theoretical idea might appear much later after the original data has already been analysed. The same problem holds for future sensitivity studies as new ideas might appear only after a sophisticated sensitivity study has already been performed. In many cases, a large amount of redundant workload would need to be dedicated if a complete restart of the data analysis procedure was performed to test such a new idea.

Fortunately, experience from current LHC studies shows that many new theories do not necessarily require the data analysis to start over from the beginning: even though the underlying physics might be very different, as long as a new model $\mathcal{M}_{\text{new}}$ predicts a collider topology which is \emph{experimentally (nearly) indistinguishable} from one which has already been analysed in the context of a different model $\mathcal{M}_{\text{old}}$, results originally determined for model $\mathcal{M}_{\text{old}}$ may be re-used to quickly derive the experimental result for model $\mathcal{M}_{\text{new}}$, \emph{without} restarting the full data analysis chain. This idea, coined as \emph{recasting}, is often used by physicists outside the experimental collaborations to confront their theoretical ideas with experimental truth. Although the used data selection criteria may not be optimal for $\mathcal{M}_{\text{new}}$, recasting often produces a sufficiently accurate result to quickly distinguish viable and excluded parameter regions.

For illustration, let us consider an example in the context of the LHC: the production of the supersymmetric partners of the neutral gauge and Higgs bosons, $\tilde \chi^0_i$, e.g.\ via the channel $p p \rightarrow \chi^0_2 \chi^0_1$ with subsequent decay $\chi^0_2 \rightarrow \chi^0_1 Z$. If the $Z$ boson decays leptonically, the final state would consist of two leptons related to the $Z$ resonance plus experimentally invisible neutralinos. A proper experimental search designed specifically for this topology would thus filter events with two high-energetic leptons with invariant mass near $M_Z$ and a large amount of missing transverse momentum, see e.g. Ref.~\cite{Aad:2014vma}. Comparing the number of observed events $O$ and the theoretically predicted numbers for both the Standard Model background $B$, its error $\Delta B$ and the signal $S \pm \Delta S$ of a particular supersymmetric spectrum, one can classify whether the observation is consistent with the model prediction or whether the model can be excluded according to a given confidence level.

Even though originally designed to search for Supersymmetry, any other model which produces a similar topology with three high energetic leptons plus further, invisible particles would predict a similar signal for this search. For example, a theory with an extended scalar sector with a pseudoscalar $A$ and a stable, neutral scalar $H$ would predicts the topology $p p \rightarrow A H, A \rightarrow Z H$ . This would experimentally look very similar to the above mentioned supersymmetric setup. Hence, we can use the same analysis strategy with same $O$ and $B \pm \Delta B$. One only needs to determine the new signal prediction $S^\prime \pm \Delta S^\prime$ which may be different from $S \pm \Delta S$ as cross sections, branching ratios or cut efficiencies due to different kinematics may be different for a different underlying physics model. In this way, a corresponding $p$-value for the Inert Doublet model can be calculated without requiring additional experimental effort as $O, B$ and $\Delta B$ can be kept, see e.g.\ Ref.~\cite{Belanger:2015kga}. Therefore, determining the new prediction $S^\prime \pm \Delta S^\prime$ and the $p$ value for a given model by using an existing analysis originally designed for a different model, is the core of the recasting procedure. 

For this idea, it is irrelevant whether the experiment has been performed already and $O$ is derived from \emph{real data} or if the discussion refers to a future experiment, may it be a high-luminosity LHC running at $\sqrt{s} = \unit[14]{TeV}$  or the ILC, for which $O$ is a \emph{fictive expected} number. As long as the same analysis procedure is used, Standard Model background and observation --- real or fictive --- can be used from an old study and re--applied on a new model easily. Therefore, sensitivity studies of future experiments performed in the context of one BSM theory can be similarly recasted into sensitivity studies for other theories beyond the Standard Model, provided that they have a common experimental signature. Recasting sensitivity studies makes best use of a given experimental analysis as it allows future theoretical ideas to re-use dedicated old studies when designing new experiments. Every new theoretical idea which can be shown to be testable at the discussed experiment yields another pro-argument to construct it. This is why recasting is useful for both the theory community, providing them with important information about expectable future results for their models of interest, and the experimental community which get additional applications of their dedicated studies for free. 

In practice, recasting an analysis is a very model independent task and requires the combination of various standard HEP software tools with only few model-dependent settings. Hence, this task can be significantly automatised and there are different tools for this purpose. In this study, we use the tool \Checkmate{} which has proven to be useful to test an arbitrary model against various results from the LHC. As it has been used by many people in the phenomenology community to determine how LHC results probe their respective model of interest, we foresee a similar popularity for people determining the ILC sensitivity to their respective models. For that purpose, we aim to extend \Checkmate{} to also take into account BSM ILC searches and discuss our current work-in-progress here.

In section \ref{sec:chaplhc}, we first summarise the principles of \Checkmate{} and how it currently determines bounds on theoretical models by using LHC results only. We discuss the required steps to satisfactorily describe ILC physics with \Checkmate{} in section \ref{sec:chapilc}. We implement an existing monophoton study as a proto-example into \Checkmate{} and use it to analyse three different dark matter benchmark models. The models and the assumed collider setup is discussed in section \ref{example} and results are shown in section \ref{sec:results}. We conclude in section \ref{sec:summary} and give an outlook over still open issues. 

\section{Testing Models at the LHC with CheckMATE}
\label{sec:chaplhc}
We first want to illustrate the steps which \texttt{CheckMATE} currently performs to test a given model at the LHC, clarifying the necessary transition steps to perform similar tasks for the ILC. 

\texttt{CheckMATE} requires an input file which provides the relevant mandatory and optional parameters. It takes as an input a BSM model that is known to the Monte Carlo (MC) event generator \texttt{MG5\_aMC@NLO} \cite{Alwall:2014hca}, an \texttt{.slha} particle spectrum file \cite{Skands:2003cj} and a set of processes to be simulated. An example for one such file is shown in Fig.~\ref{fig:example} which simulates 10,000 Monte-Carlo events for one process, here gluino pair production in Supersymmetry.

\begin{figure}
\centering
\includegraphics[width=\columnwidth]{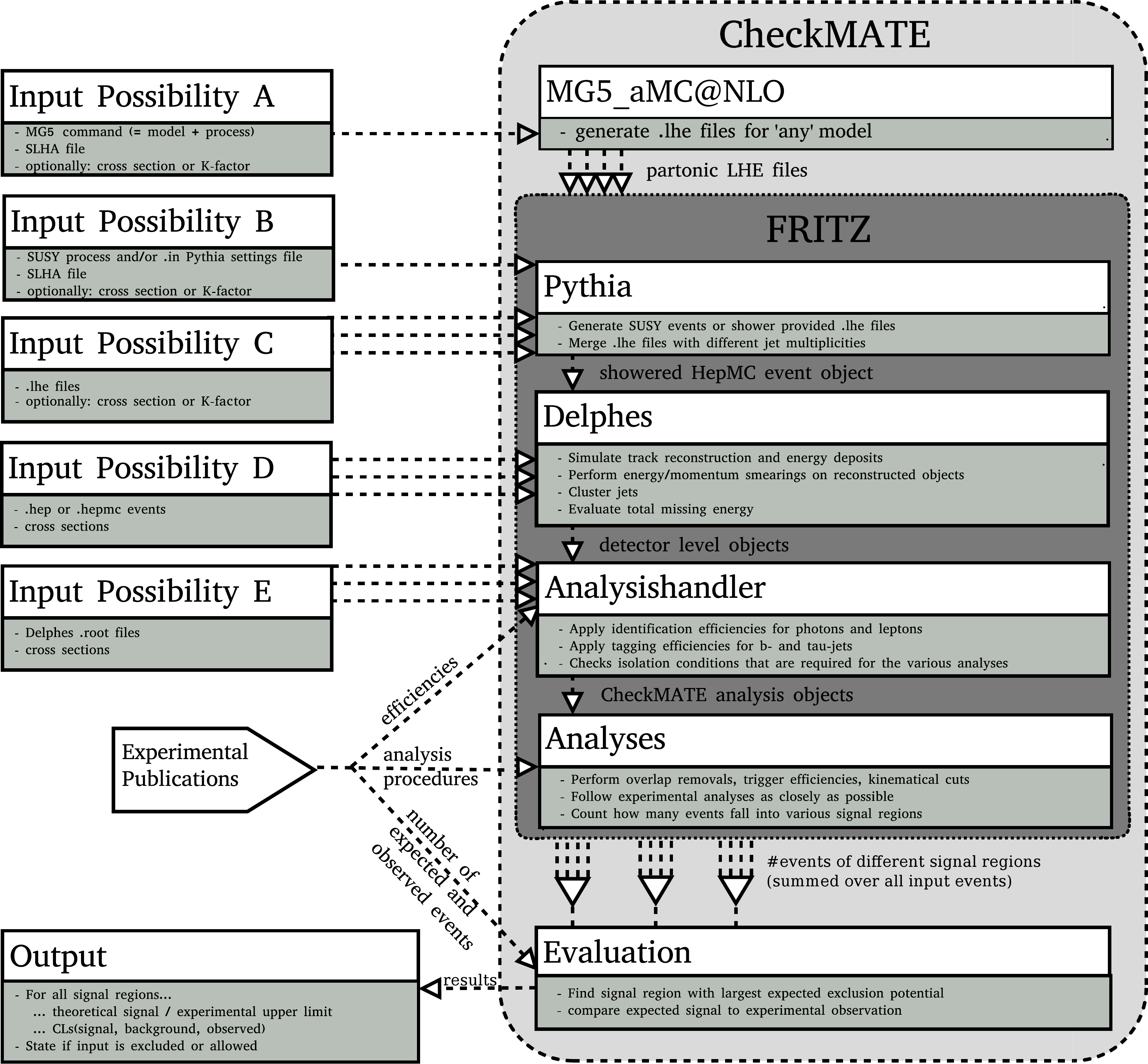}
\caption{Flowchart to illustrate which steps \texttt{CheckMATE} performs to test a given input by the user.}
\label{fig:oldflow}
\end{figure}

The following chain of tasks is then performed completely automatically and an illustrative flowchart is provided in Fig.~\ref{fig:oldflow}. First, \texttt{CheckMATE} calls the event generator \texttt{MG5\_aMC@NLO} to simulate the partonic events of the given process. Internally, \texttt{MG5\_aMC@NLO} derives the matrix element for in principle any BSM theory from the vertex information stored in the respective \texttt{UFO} \cite{Degrande:2011ua} file which may either be created by using the model building tools \texttt{FeynRules}~\cite{Christensen:2008py,Alloul:2013bka} or
\texttt{SARAH}~\cite{Staub:2013tta} or can be downloaded from the \texttt{FeynRules} website\footnote{\url{http://feynrules.irmp.ucl.ac.be/}} for some popular models.
After the parton events are generated, the hadron shower \texttt{Pythia8} \cite{Sjostrand:2007gs} is called to translate these into fully hadronic MC events as they could have taken place at the LHC if the input model was true. These events are then automatically passed through the fast detector simulation \texttt{Delphes} \cite{deFavereau:2013fsa} which simulates \texttt{ATLAS} and \texttt{CMS}, the two multipurpose detectors at the LHC. The results of this simplistic fast detector simulation are further refined by applying additional, $p_T$ and $\eta$ dependent efficiency functions which describe the probability to reconstruct particular final state objects. The output of this simulation is then analysed by a \texttt{CheckMATE}-internal software framework which applies the same event selection procedure as the experimental collaborations quote in their publications. The code then finally tests the compatibility of the prediction derived from the user's input to the experimental observation taken from the respective publication. If this discrepancy is larger than the 95\% C.L., \texttt{CheckMATE} returns ``\emph{excluded}'', otherwise the response is ``\emph{allowed}''. 

For more information we refer to the \Checkmate{} manuals in Refs.~\cite{Drees:2013wra,Kim:2015wza,Dercks:2016npn}.

\begin{figure}
\begin{Verbatim}[commandchars=\\\@\@,frame=lines,fontsize=\footnotesize]
[Parameters]
SLHAFile: /scratch/benchmark1.slha

[squ_asq]
MG5Process: import model mssm; 
            generate p p > go go
MaxEvents: 10000
\end{Verbatim}
\caption{Minimal working example for an input file to test a supersymmetric parameter point in \texttt{CheckMATE} in the gluino pair production channel.}
\label{fig:example}
\end{figure} 

\section{Testing ILC Physics with CheckMATE --- A To-Do List}
\label{sec:chapilc}
\texttt{CheckMATE} is comfortable to use since it performs the necessary steps for simulation, data analysis and statistical evaluation completely automatically. For that purpose, \Checkmate{} connects to a variety of tools, most importantly \texttt{MG5\_aMC@NLO}, \texttt{Pythia8} and \texttt{Delphes}. As we aim to keep this automatisation, we need to determine how to adapt these programs when changing from the Large Hadron Collider to the International Linear Collider. Furthermore, contrarily to the LHC which has a fixed experimental design, the ILC currently is in a planning stage and some experimental specifications need to be accessible as free parameters to allow the users to study how sensitivities change when changing these experimental parameters. 

\subsection{Defining the Collider Settings}
As stated above, in contrast to the LHC which is a running experiment with fixed parameters, some important collider parameters of the ILC are still in discussion. The sensitivity of this future experiment to a given BSM hypothesis may depend on the details of these parameters. Therefore, they need to be accessible by the users as free parameters, also to be able to discuss how sensitivites may change with a different experimental setup. The following properties have to be manipulable:
\begin{description}
\item[Centre-of-mass energy $\sqrt{s}$] As the specific staged scenario is still under discussion, different assumptions may be made about the centre-of-mass energy of the ILC. Clearly, as this strongly affects cross sections and kinematic distributions, different assumptions for $\sqrt{s}$ will change the prediction for both signal and background. 
Ideally, we would like $\sqrt{s}$ to be a free parameter the user can choose at will. However, in practice this is a goal which is very complicated to achieve: as explained above, the advantage of the recasting procedure is that event selection and Standard Model background numbers $B \pm \Delta B$ can be kept and only the new prediction $S \pm \Delta S$ for the new signal model needs to be determined. Not only do the numbers $B \pm \Delta B$ change upon changing $\sqrt{s}$ but also the analysis procedure itself may be affected: higher values of $\sqrt{s}$ typically require stronger cuts on the final state objects to reduce the contamination from Standard Model background. Both the analysis and the resulting background prediction would need to be re-determined if the user chooses an arbitrary value of $\sqrt{s}$ which currently is not manageable.

Therefore we follow a similar strategy as \Checkmate{} currently uses for the LHC. Here, there often exist similar versions of the same analysis at both $\sqrt{s} = 8$ and $\unit[13]{TeV}$, however with slightly different cuts and different background numbers. Similarly, for the ILC each implemented analysis assumes one specific value of $\sqrt{s}$ --- in our study $\sqrt{s} = \unit[500]{GeV}$ --- and with this value fixed, the background prediction $B \pm \Delta B$ is constant. However, the same analysis may then be re-implemented for a different value, e.g.\ $\sqrt{s} = \unit[1]{TeV}$ with a new prediction $B^\prime \pm \Delta B^\prime$. Users then may choose for which of the fixed, implemented values of $\sqrt{s}$ they want to do the model test. For this study, only one value of $\sqrt{s} = \unit[500]{GeV}$ has been implemented. 

\item[Lepton Polarisations $P_{e^+}, P_{e^-}$] Another very important, ILC-specific aspect is the possibility to set the spin polarisation of the initial state leptons. Such a feature would greatly enhance the discovery potential for some new physics models due to signal enhancement and background suppression. Moreover, once a signal has been found, analysing the dependence to the initial state spins would help understanding the chiral details of the underlying physics. We therefore allow the user to define the polarisation $P_{e^+}, P_{e^-}$ of the positron and electron as respective free parameters. The signal is then generated with that specific polarisation set. However, also the background prediction $B \pm \Delta B$ depends on the polarisation. Fortunately this dependence is not arbitrary but can be formulated in general as
\begin{align}
B(P_{e^+}, P_{e^-}) = \frac{1}{4}\Big(& (1+P_{e^-})(1+P_{e^+}) B(+1, +1) \nonumber \\
+ & (1+P_{e^-})(1-P_{e^+}) B(+1, -1)  \nonumber \\
+ & (1-P_{e^-})(1+P_{e^+}) B(-1, +1)  \nonumber \\
+ & (1-P_{e^-})(1-P_{e^+}) B(-1, -1) \Big). \label{eq:bkgpol}
\end{align}
Here, $B(\pm 1, \pm 1)$ is the background prediction which corresponds to a degree of polarisation $P_{e^+} = \pm \unit[100]{\%}, P_{e^-} = \pm \unit[100]{\%}$ And $+$ ($-$) denotes right- (left-) handed polarisation. An analogous formula can be defined for $\Delta B$.  Hence, contrarily to the LHC for which we only needed one pair $B \pm \Delta B$, we need to provide four sets of numbers for each ILC analysis, i.e. each $B(\pm 1, \pm 1)$ and the respective numbers $\Delta B (\pm1, \pm 1)$. Whenever the user chooses a nontrivial lepton polarisation $(P_{e^+}, P_{e^-})$, \Checkmate{} uses Eq.~(\ref{eq:bkgpol}) to determine the corresponding background numbers $B(P_{e^+}, P_{e^-}) \pm \Delta B (P_{e^+}, P_{e^-})$. 

\item[Integrated Luminosity $\lint$] When performing a future sensitivity study, one needs to fix the value for the  integrated luminosity that is assumed in order to state after which time period the given result could be accomplished. For example, in this study, we use $\lint = \unit[500]{fb}^{-1}$ which are expected after the initialisation phase of the first $3$ to $5$ years and on a yearly basis afterwards \cite{Brau:2016cym}. Users should be able to set this parameter freely with \Checkmate{} scaling signal and background numbers accordingly. \Checkmate{} uses the background numbers stored for a base value of \unit[1]{fb}$^{-1}$ and rescales them to the user's target luminosity by simply multiplying $B$ by $\lint$. For the error $\Delta B$, we currently conservatively assume that these are of systematic nature and therefore equally scale with $\lint$. It is forseen to extend this treatment and to allow a splitting of the error into a systematic and a statistic component, with the former scaling with $\lint$ and the latter with $\sqrt{\lint}$, respectively. 

\item[Collider scenarios] The reason to make the above mentioned parameters accessible by the user is to allow systematic studies of how the sensitivity depends on certain assumptions. If we want to compare these assumptions, it would be comfortable if we can test all possible combinations simultaneously with \Checkmate. In a related context, if we use polarised beams and assume an overall time scale for the ILC, we can think of different possibilities how long the ILC will respectively run in the different polarisation modes $++, +-, -+$ and $--$. Again, if we split the total integrated luminosity into four separate runs with individual lepton polarisations it would be convenient if \Checkmate{} simulates all these simultaneously and determines the strongest bound which can be derived from all individual results.

Within the version that we used for this study, we allow the user to define various \emph{collider scenarios} with different combinations of $\lint, P_{e^+}$ and $P_{e^-}$. Our study, in particular, tests the following scenarios:
\begin{itemize}
\item \unit[500]{fb}$^{-1}$ with unpolarised beams,
\item \unit[500]{fb}$^{-1}$ with polarised beams, split into
\begin{itemize}
\item[$\circ$] \unit[200]{fb}$^{-1}$, $P_{e^-} = +\unit[80]{\%}$, $P_{e^+} = -\unit[30]{\%}$,
\item[$\circ$] \unit[200]{fb}$^{-1}$, $P_{e^-} = -\unit[80]{\%}$, $P_{e^+} = +\unit[30]{\%}$,
\item[$\circ$] \color{white}0\color{black}\unit[50]{fb}$^{-1}$,  $P_{e^-} = +\unit[80]{\%}$, $P_{e^+} = +\unit[30]{\%}$,
\item[$\circ$] \color{white}0\color{black}\unit[50]{fb}$^{-1}$,  $P_{e^-} = -\unit[80]{\%}$, $P_{e^+} = -\unit[30]{\%}$
\end{itemize}
\end{itemize}
\Checkmate{} then runs each of these separately and considers each result as an individual measurement. On longer terms, we aim to combine the individual measurements automatically to determine a much stronger total result.
\end{description}

\subsection{Event Generation}
Whilst \texttt{MG5\_aMC@NLO} is a very powerful tool to simulate partonic events for a hadron collider for nearly any theory beyond the Standard Model, it lacks the proper description of beam effects which are relevant for a high energy linear collider. Most importantly, it can neither account for a proper description of initial state radiation nor for an energy spread in the centre-of-mass energy due to beam-beam interactions.

Both of these features can be properly simulated by the Monte-Carlo event generator \texttt{Whizard} \cite{Kilian:2007gr,Moretti:2001zz,Nejad:2014sqa,Kilian:2011ka}, a Monte Carlo generator which is also capable of simulating events for BSM theories --- a mandatory feature to be useful for a tool like \texttt{CheckMATE}. One necessary step is therefore to link \texttt{CheckMATE} to \texttt{Whizard} such that it can be called in an analogous way as can currently been done with \texttt{MG5\_aMC@NLO}, see Fig.~\ref{fig:example}. Showering and hadronisation of the event are still performed by processing the partonic events with \texttt{Pythia8}.

\subsection{Detector Simulation}
For a proper description of experiments at particle colliders it is not sufficient to only simulate the collision itself. One also needs to take into account that a realistic detector may not observe the true final state but only registers visible objects in a finite coverage area and reconstructs their energies and momenta including intrinsic systematic uncertainties. A full consideration of all these effects would require the simulation of each particle's trajectory in the detector. Whilst this is in principle possible, e.g.\ via \texttt{Geant4} \cite{ALLISON2016186}, such a simulation typically requires days to process an entire Monte-Carlo event sample  and thus becomes very unfeasible for a tool like \Checkmate{} which normally needs to analyse hundreds or thousands of model points in a given theory framework.

For that purpose, \Checkmate{} uses the fast detector simulation \texttt{Delphes} which applies efficiency functions to account for the finite detection probabilities and reconstruction uncertainties. As it is well connected to all the other modules in \Checkmate{}, we try to use the same tool to describe the ILD detector. Fortunately, \texttt{Delphes} already provides a standard description of this detector based on ILC Technical Design Report, Ref.~\cite{Behnke:2013lya}. Though giving a good first order estimate of the relevant coverages and efficiencies, there are a couple of drawbacks worth mentioning:
\begin{itemize}
\item The efficiencies have been taken from the technical design report published in 2013 and need to be updated occasionally with respect to recent layouts decisions.
\item Only the hadronic calorimeter (HCal) and the electromagnetic calorimeter (BCal) are taken into account. It is especially the BCal which is not simulated within \texttt{Delphes} and therefore the forward detector region is not properly described. 
\item \texttt{Delphes} only translates particles which appear in the original Monte Carlo sample and does not account for additionally reconstructed final state objects, for example produced by secondary beam interactions $ \gamma \gamma \rightarrow \text{hadrons}$. 
\end{itemize}
As we see later, a proper BSM event selection is designed in such a way that no signal events are expected for which the above effects are relevant. Therefore, they do not play a significant role in the determination of the signal prediction. However, they are very important for the consideration of Standard Model background contaminations and therefore need to be taken into account in the long term to be able to use \Checkmate{} not only to produce $S \pm \Delta S$ but also $B \pm \Delta B$. At its current stage, we rely on experimental studies performed with a full detector simulation which give us $B \pm \Delta B$.\footnote{Note that for LHC searches, the same strategy is pursued: \Checkmate{} takes $O$ and $B \pm \Delta B$ from the experimental publications and only re-determines $S \pm \Delta S$ for the user's model point.}

\subsection{Analyses}
\label{subsec:analys}
\Checkmate's arguably most important module  is the analysis framework. In here, \Checkmate{} determines if the input topology provided by the user produces a final state which would have passed the selection criteria of a recasted BSM search. One powerful aspect of \Checkmate{} are the many implemented LHC analyses, i.e.\ 30 at $\sqrt{s} = \unit[8]{TeV}$ and 19 at $\sqrt{s} = \unit[13]{TeV}$.  These cover various different topologies and there is a large probability that an arbitrary topology provided by the user is covered by any of these. To achieve a similar effectiveness in the context of ILC searches, we aspire a large number of implemented topologies on the long term as well. 

For the beginning, we start with one arguably simple analysis strategy as a proof-of-principle, based on the selectio described in Ref.~\cite{Habermehl:2017dxh}. This analysis, typically called \emph{mono-photon} search, is motivated by any theory with a weakly interacting massive particle (WIMP) $\chi$ as a dark matter candidate. If such a model predicts any 4-particle diagram $e^+ e^- \bar \chi \chi$, a lepton collider is expected to produce events $e^+ e^- \rightarrow \bar \chi \chi \gamma$ with a photon either coming from initial state radiation or, depending on the model, from the physics inside the $ee\chi\chi$ interaction. The final state then produces an event with a single, high-energetic photon and missing energy due to the momentum inbalance produced by the invisibly escaping dark matter particles, see also Refs.~\cite{Bartels:2012ex, Dreiner:2012xm, Habermehl:2017dxh}.  Our analysis is defined by the following event selection procedure:
\begin{enumerate}
\item At least one photon needs to have 
\begin{itemize}
\item $E \leq \unit[220]{GeV}$, 
\item $|\cos \theta | < 0.996$ and
\item $p_T > \unit[5.71]{GeV}$ if $|\phi| \leq 0.67195$, else $p_T > \unit[1.97]{GeV}$,
\end{itemize}
Out of all photons which pass these constraints, the one with the highest energy is called signal photon.
\item No track with $p_T > \unit[3]{GeV}$ must be reconstructed.
\item The sum of energies of all visible objects minus the energy of the signal photon must not exceed \unit[20]{GeV}.
\item No energy deposit must be registred within the \texttt{BCal}.
\end{enumerate}
The first cut vetos events without the necessary signal photon and rejects many photons from the irreducible background process $e^+ e^- \rightarrow \bar \nu \nu \gamma$ whose $Z$-resonance peak appears at $E^\text{max}_\gamma = (\sqrt{s} - M_Z^2/\sqrt{s})/2 \approx \unit[240]{GeV}$.  As no charged particle is expected in the signal final state $\chi \bar \chi \gamma$, a veto on charged particles is applied. The third cut vetoes events with more than one photon, most importantly the Standard Model process $e^+ e^- \rightarrow \gamma \gamma$. Lastly, a veto on particles in the forward region rejects Standard Model Bhabha events $e^+ e^- \rightarrow e^+ e^- \gamma$ with an ISR photon and a small scattering angle.\footnote{Note that at its current stage, the last cut has an efficiency of \unit[100]{\%} as \texttt{Delphes} does not consider the forward detector region. However, a typical dark matter signal would not be affected by this cut and therefore the signal prediction should not be significantly affected by this incomplete detector description.} For more details about this analysis we refer to Ref.~\cite{Habermehl:2017dxh}.

As stated before, we need to implement background predictions $B \pm \Delta B$ for each of the four basis polarisation settings $P_{e^+} = \pm 1, P_{e^-} = \pm 1$ for $\lint = \int \mathcal L = \unit[1]{fb}^{-1}$. For the above analysis, these are as follows:\footnote{These numbers have been determined by M. Habermehl with the same setup as described in Ref.~\cite{Habermehl:2017dxh}, using the updated event selection criteria described here.}
\begin{align}
P_{e^-} = +1, P_{e^+} = +1:&& 208.85 &\pm 1.19, && && && && \\
P_{e^-} = +1, P_{e^+} = -1: && 707.46 &\pm 4.03, && && && && \\
P_{e^-} = -1, P_{e^+} = +1:&& 13759.66 &\pm 78.35,&& && && &&\\
P_{e^-} = -1, P_{e^+} = -1:&& 208.85 &\pm 1.19.&& && && &&
\end{align}
(Note that the largest background comes from Standard Model neutrino interactions which originate from  left-chiral interactions. For a respective left-handed polarisation, the background contamination is very large.

\section{Validation and Recasting}
\label{example}
We now want to test the above features for some example cases. For that purpose, we define three benchmark models with different couplings between the Dark Matter and the Standard Model sector. Most importantly, one of these models has also been analysed in Ref.~\cite{Habermehl:2017dxh} using a full detector simulation and we compare our corresponding bounds\footnote{The authors of Ref.~\cite{Habermehl:2017dxh} provided us with their updated signal predictions using the event selection described in this study.} to show how well  \Checkmate{} reproduces the result from the full experimental analysis. Practical information about the extended \Checkmate{} input card used for this analysis is provided in App.~\ref{sec:setup}.

\subsection{Benchmark Models}
\begin{description}
\item[Effective Vector-Like Interaction]
The model analysed in Ref.~\cite{Habermehl:2017dxh} considers a simplified dark matter model which assumes a high mass particle with mass $m_{\text{med}} \gg \sqrt{s}$ mediating the interaction of the fermionic WIMP candidate $\chi$ and the Standard Model leptons. This can be formulated as an effective interaction
\begin{align}
\mathcal{L}_1 \supset - \frac{g_\chi g_f}{m_{\text{mediator}}^2} (\bar f \gamma^\mu f)(\bar \chi \gamma_\mu \chi). \label{eq:eff1}
\end{align}
In this case, the photon in the process $e^+ e^- \rightarrow \bar \chi \chi \gamma$ originates from initial state radiation of either of the two leptons. The appealing feature of such an effective Dark Matter models is that it only depends on two parameters, i.e.\ the mass of the WIMP candidate $\chi$ and $\sqrt{g_\chi g_f}/m_{\text{mediator}}$. In the following, we fix $g_f = g_\chi =1$ and use the mass $m_{med}$ as a free parameter. Bounds can trivially be rescaled for cases with non-unit $g$.

\item[Simplified Scalar Interaction]
The vector-like interaction shown in Eq.~(\ref{eq:eff1}) typically originates from a UV-complete theory with a spin-1 mediator particle. In our second model, we change the mediator to a scalar particle and remove the requirement $m_{\text{med}} \gg \sqrt{s}$, i.e. the mediator can have a light mass. The interaction hence looks like
\begin{align}
 \mathcal{L}_2 \supset - m_{\text{med}}^2 |\phi|^2 - [g_f \phi (\bar f f) \chi) + g_\chi \phi (\bar \chi \chi) + \text{h.c.}] \label{eq:eff2}
\end{align}
The expected topology is the same as in the above effective vector model. However whilst for the LHC the spin-dependence of the interactions in Eqs.~(\ref{eq:eff1}) and (\ref{eq:eff2}) is barely noticable, it leads to significant differences at the ILC as here spin polarisation of the initial state leptons may increase or decrease the production cross section depending on the type of interaction. Also, in regions for which the mediator mass is of the order of $\sqrt{s}$ or below the bound is expected to change compared to the effective case as the mediator can be produced on-shell. Note that for the sake of simplicity we again fix $g_f = g_\chi = 1$ and the total width $\Gamma_\phi$ of the scalar particle to be \unit[1]{GeV}.\footnote{In a proper, UV complete theory the width $\Gamma_\phi$ would depend on the details of the dark matter sector, including the parameters $g$ and $m_\chi$}

\item[Photino-like Dark Matter]
We define a third model which uses a dark matter candidate $\chi$ and two scalars $\tilde f_L$, $\tilde f_R$ which couple to electrons via the following interaction:
\begin{align}
\mathcal{L}_3 \supset -e \tilde f^*_L (\bar \chi P_L f) - e \tilde f_R^* (\bar \chi P_R f) + \text{h.c.}
\end{align}
This corresponds to a supersymmetric scenario in which the lightest neutralino is a pure photino. In this toy model, by changing the mass of the particles $\tilde f_L, \tilde f_R$ we can change how $\chi$ couples to left- and right-handed leptons, respectively. We therefore choose $m_{\tilde f_L}, m_{\tilde f_R}$ as the free parameters of this model and fix $m_\chi = 0$. 
\end{description}
We implemented these models in \texttt{FeynRules} and exported them to the event generator \texttt{Whizard} to make them accessible in \Checkmate.

\begin{figure}
\centering
\includegraphics[width=\columnwidth]{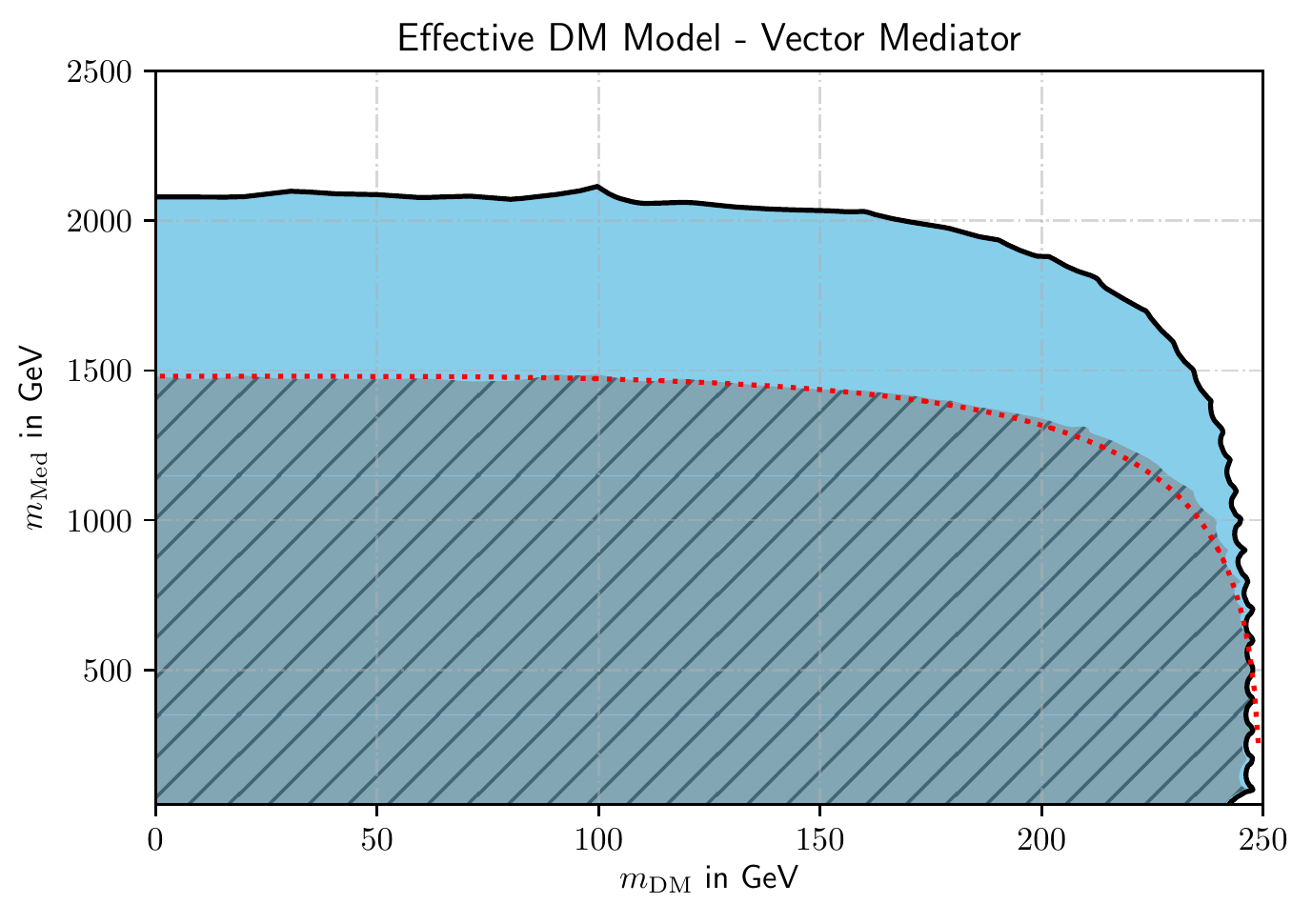}
\caption{Bounds on the effective vector dark matter model, see Eq.~\ref{eq:eff1} determined with \Checkmate{}. Hatched regions show the result using unpolarised beams, coloured regions the results with polarised beams (see text). The red dotted curve shows the result for unpolarised beams determined by the authors of Ref.~\cite{Habermehl:2017dxh} using a full detector simulation. Wave-like features in the contour are artifacts of the interpolation procedure and have no physical meaning.}
\label{fig:vector}
\end{figure}

\begin{figure}
\centering
\includegraphics[width=\columnwidth]{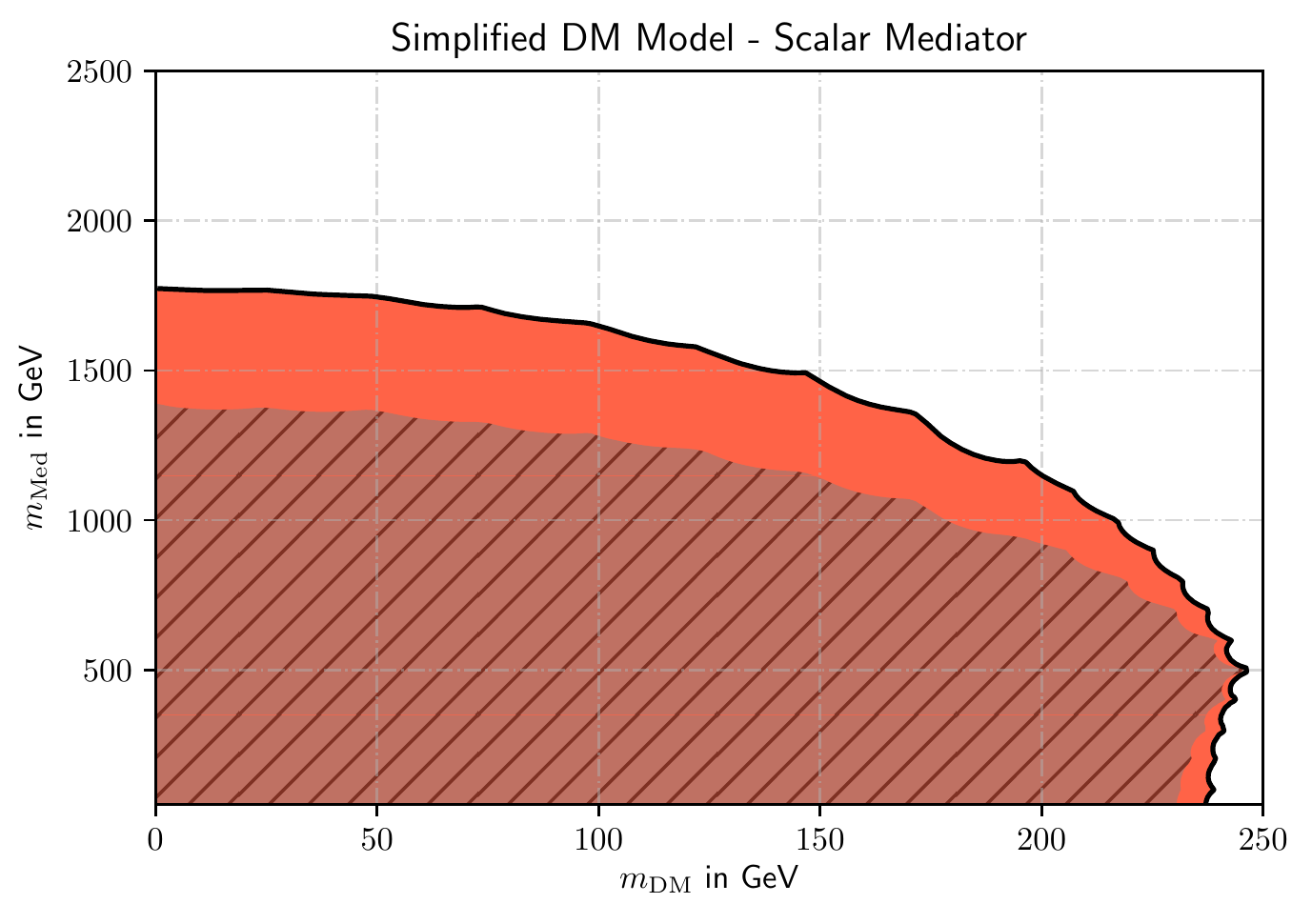}
\caption{Bounds on the simplified dark matter model, see Eq.~\ref{eq:eff2}. Information as in Fig.~\ref{fig:vector} but without the full detector simulation result.}
\label{fig:scalar}
\end{figure}

\begin{figure}
\centering
\includegraphics[width=\columnwidth]{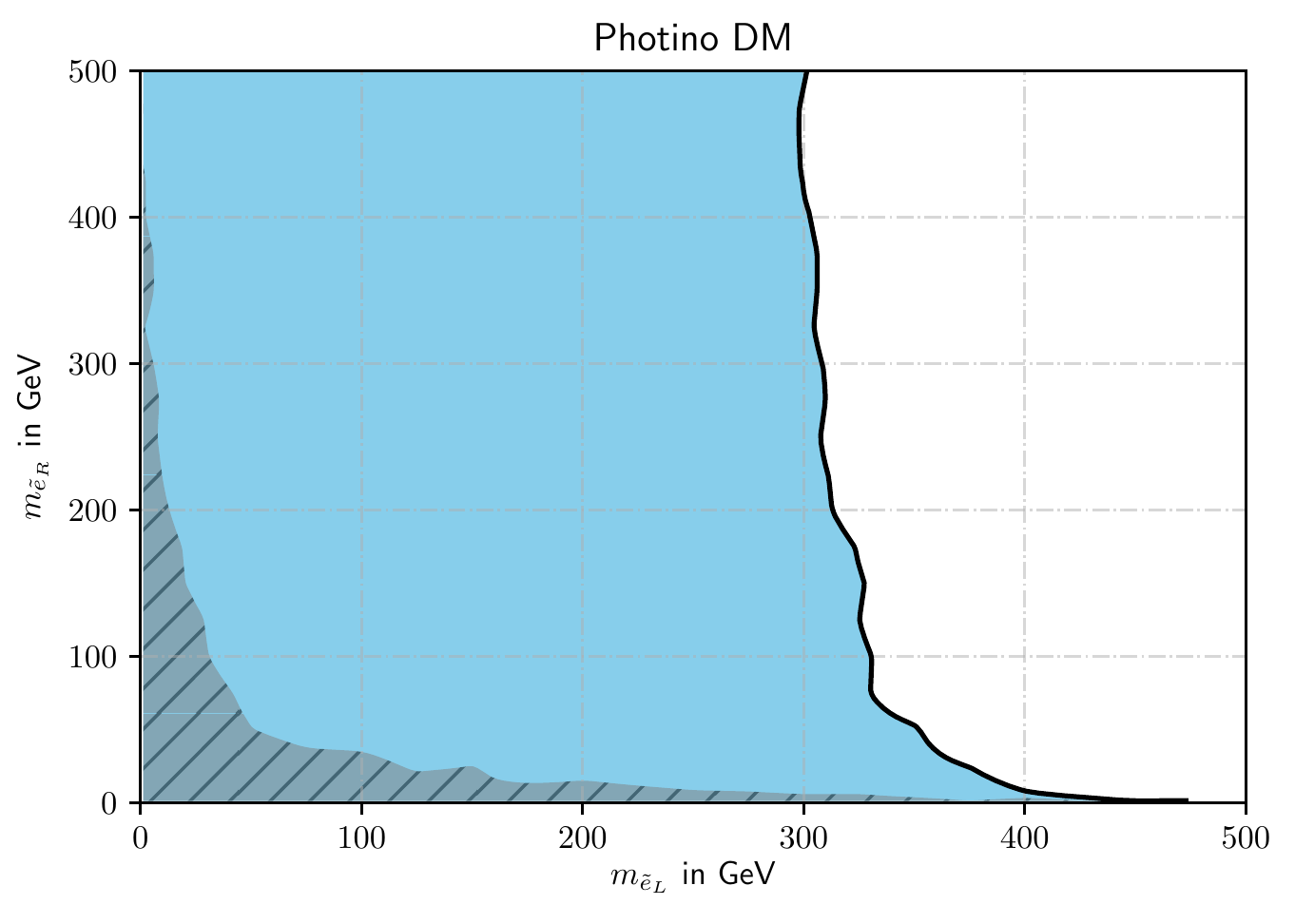}
\caption{Bounds on the photino-like dark matter model, see Eq.~\ref{eq:eff2}. Information as in Fig.~\ref{fig:vector} but without the full detector simulation result.}
\label{fig:photino}
\end{figure}

\section{Results}
\label{sec:results}
We now discuss the preliminary results determined with \Checkmate{}. These are shown in Figs.~\ref{fig:vector}-\ref{fig:photino} for the three benchmark models defined before. Each plot shows the excluded area in the plane spanned by the two free mass parameters of the model. In all cases, the $//$-hatched grey area  shows the region which could be excluded at $\sqrt{s} = \unit[500]{GeV}$ with $\lint = $ \unit[500]{fb}$^{-1}$ using \emph{unpolarised} beams. In comparision, regions marked in colour could be excluded if the ILC runs with \emph{polarised} beams. As stated above, \Checkmate{} considers each measurement independently and determines which of the different polarisation phases  produced the strongest bound: light blue coloured regions are best covered with \unit[200]{fb}$^{-1}$ of data using the initial state polarisation $P_{e^-} = +\unit[80]{\%}, P_{e^+} = -\unit[30]{\%}$. Alternatively, red coloured regions prefer \unit[50]{fb}$^{-1}$ data with polarisation $P_{e^-} = +\unit[80]{\%}, P_{e^+} = +\unit[30]{\%}$, respectively. Note that the other two polarisation settings are tested but they never produce stronger results than those determined from the above two runs which is why there is no colour coding for these. All bounds correspond to an exclusion at the \unit[95]{\%} confidence level.

\subsection{Effective Vector-Like Interaction}
We start with an effective WIMP model with a vector-like interaction, see Fig.~\ref{fig:vector}. Here we observe that an unpolarised ILC setting with $\sqrt{s} = \unit[500]{GeV}$ would be able to exclude mediator masses up to \unit[1.5]{TeV} for dark matter lighter than \unit[150]{GeV}. Above this mass threshold, kinematic effects reduce the total cross section and gradually weaken the bound on $m_{\text{med}}$ up to the kinematic threshold of $m_{\text{DM}} = \sqrt{s}/2 = \unit[250]{GeV}$. If the ILC runs with a polarised initial state, the bound becomes stronger, $m_{\text{med}} \gtrsim \unit[2.1]{TeV}$ for light Dark Matter, which is explained by  a significant background suppression and a cross section enhancemenet of the signal: the electroweak Standard Model background $\nu \nu \gamma$ is mostly suppressed for right-chiral particles which is why the setup with a right-handed electron polarisation of $P_{e^-} = +\unit[80]{\%}$ results in the strongest bound. Moreover, due to the vector-like signal interaction, the model prefers an alignment of the two spins and thus favours $P_{e^+} = -\unit[30]{\%}$. All in all, we see that initial state polarisation can improve the bound on the mediator scale by nearly \unit[40]{\%}, even though only a fraction (\unit[200]{fb}$^{-1}$) of the full dataset (\unit[500]{fb}$^{-1}$) uses the prefered initial state polarisation.  In the same plot, we show the corresponding result for this model determined from a full detector simulation by the authors of  Ref.~\cite{Habermehl:2017dxh}. We observe a very good agreement and therefore are confident that our fast detector simulation produces acceptable results. Also, this renders the upcoming, recasted results more reliable. 

Note that, for comparison, the most recent ATLAS LHC measurement of the interaction $\bar q q \bar \chi \chi$, see Ref.~\cite{Aaboud:2017phn}, excludes mediator masses $m_{\text{med}} < \unit[3.1]{TeV}$ for light WIMP masses. Note, however, that this result uses two years of data and the effective operators $\bar q q \bar \chi \chi$ and $e^+ e^- \bar \chi \chi$ need not necessarily be related. For example, a leptophilic WIMP which only couples to Standard Model leptons could only be probed by the ILC.

\subsection{Simplified Scalar Interaction}
We continue with a discussion of the results shown in Fig.~\ref{fig:scalar}, determined for a signal model with a scalar mediator which is not necessarily off-shell. For unpolarised beams, we again observe the strongest bound $m_{\text{med}} \gtrsim \unit[1.4]{TeV}$ in the limit of small WIMP masses, $m_\chi \rightarrow 0$. This bound again weakens for higher DM masses due to kinematic suppression of the cross section. Note that the suppression with increasing $m_\chi$ behaves differently than in the previous scenario: a different angular momentum dependence of the interaction (scalar instead of vector) changes the dependence of the final state particles' masses, even in the case with no spin polarisation. Moreover, as we do not consider an effective theory, we approach the resonance peak in the parameter region $m_{\text{med}} \approx \sqrt{s} = \unit[500]{GeV}$ which also affects the final result. Especially, for $m_{\text{med}} \approx \unit[500]{GeV}$ we observe the strongest bound on $m_\chi$ which is weakened for both smaller and larger values of $m_{\text{med}}$. Note that this illustrates an important issue of effective field theories like the one discussed before: they are by construction only meaningful in regions which fulfill the effective approximation $m_{\text{med}} \gg \sqrt{s}$. In our analysis, this assumption is broken for $m_{\text{med}} \lesssim \unit[500]{GeV}$ and we can see that the results indeed change when going from the effective to the on-shell mediator picture, see also e.g.\ Ref.~\cite{Busoni:2013lha} 

Similarly to before, we also show the results for the case of polarised beams and we again obsere a significantly stronger bound, $m_{\text{med}} \gtrsim  \unit[1.75]{TeV}$ in the limit $m_{\chi} \rightarrow 0$. This \unit[25]{\%} stronger bound may be achieved from a scenario which uses \unit[50]{fb}$^{-1}$ of data taken with $P_{e^-} = 0.8, P_{e^+} = +0.3$. Here, the background suppression still prefers a right-handed electron polarisation. However, for an enhancement of an interaction mediated by a \emph{scalar} particle, the spins prefer anti-alignment and thus the strongest bound is achieved for $P_{e^+} = +0.3$. It is remarkable \unit[50]{fb}$^{-1}$ of data using polarised beams is sufficient to produce a much stronger bound than \unit[500]{fb}$^{-1}$ of data, i.e.\ a ten times larger statistic, recorded with unpolarised beams. Furthermore, note that even though both this and the previous scenario prefered different lepton polarisations, the sensitivty to \emph{both} scenarios can be increased simultaneously if the total amount of \unit[500]{fb}$^{-1}$ achievable within one year is divided into the four polarised subruns.

\subsection{Photino-Like Dark Matter}
Lastly, we discuss the results of our third toy scenario  which couples a photino-like DM candidate to Standard Model partcles via two scalars $\tilde f_L, \tilde f_R$ that respectively couple to left-handed or right-handed leptons only. Our results are shown in Fig.~\ref{fig:photino}. When running in unpolarised mode, no chirality is prefered and thus the bound is symmetric in the $m_{\tilde f_L}$-$m_{\tilde f_R}$-plane. In case of degeneracy, $m_{\tilde f_L} = m_{\tilde f_R} =: m_{\tilde f}$ the mass bound excludes $m_{\tilde f} \lesssim \unit[50]{GeV}$. The mass bound for one scalar may be weakened if the mass of the respective other scalar is increased accordingly. For $m_{\tilde f_{L/R}} \gtrsim \unit[400]{GeV}$ the respective other particle $\tilde f_{R/L}$ may even be massless. 

In comparison, using polarised beams can significantly enhance the result. Here, an absolute mass bound of $m_{\tilde f_R} \gtrsim \unit[300]{GeV}$ can be achieved which is independent of the mass of $m_{\tilde f_L}$. The bound can increase to $m_{\tilde f_R} \gtrsim \unit[400]{GeV}$ if $\tilde f_L$ is lighter than \unit[10]{GeV}. These bounds could be derived from \unit[200]{fb}$^{-1}$ of data taken at polarisation $P_{e^-} = +\unit[80]{\%}, P_{e^+} = -\unit[30]{\%}$ which has the largest background suppression rate. 

One should note that charged scalars in such a mass range would be more precisely analysable in direct searches and our results only provide complementary information from a different channel. It mostly illustrates \Checkmate{} can conveniently derive the ILC sensitivity to a given model with nontrivial chiral structure by using different assumptions about the polarisation of the initial state leptons.

\section{Summary}
\label{sec:summary}
Recasting is a powerful procedure which applies existing collider results on new theoretical ideas without requiring the full experimental data analysis to be restarted. Theories which share experimentally indistinguishable topologies can be tested via identical event selection techniques, bearing the advantage that background expectation and the number of observed events stay constant. \Checkmate{} is a powerful tool which uses Monte Carlo recasting techniques to  test in principle arbitrary BSM theories against a large amount of existing LHC results from both the $\sqrt{s} = 8$ and $\unit[13]{TeV}$ runs. We aim to extend this functionality to allow future sensitivity studies for the ILC to be recasted in a similar manner, hopefully allowing phenomenologists to conveniently produce more use-cases for the ILC as a discovery machine for physics beyond the Standard Model.

In this work we presented our current work in progress to achieve this goal. We were able to make use of the powerful features of the Monte Carlo event generator \texttt{Whizard} to be able to simulate $e^+ e^-$ collisions including the effects of beam polarisation, initial state radiation and beamstrahlung. Furthermore, we continue using the fast detector simulation \texttt{Delphes} to describe the ILD detector and it currently provides a good approximation of the most relevant acceptance and efficiency factors. Lastly, we extended the set of accessible parameters in \Checkmate{}, allowing users to test different polarisation and luminosity combinations. This makes it very convenient to discuss the importance of e.g.\ the lepton polarisation for the overall sensitivity of the experiment to a given BSM hypothesis.

Many theories with a dark matter candidate produce mono-X topologies at a collider experiment and we implemented a monophoton analysis in \Checkmate{} which aims to record the topology $e^+ e^- \rightarrow \chi \chi \gamma$. To test the quality of our approximations, most importantly the fast detector simulation, we implemented an analysis similar to the event selection given in Refs.~\cite{Bartels:2012ex,Dreiner:2012xm,Habermehl:2017dxh} and tested the same toy model, a fermionic WIMP with an effective vector-like interaction between the dark matter and the Standard Model sector. To illustrate the recasting aspect of \Checkmate{}, we analysed two more benchmark models with slightly different assumptions about the interaction between the WIMP candidate $\chi$ and the initial state leptons. We observe good agreement of our bounds with the result determined with the full detector simulation and conveniently determined corresponding limits in the other two benchmark scenarios. In all cases, we observe a significant improvement of the bounds for all models within the same overall data taking time scale if the initial state leptons are polarised.

Lastly, we eventually intend to publish the \Checkmate{} version which we used to perform the studies performed in this work. To do this, some parts of the program still need to be improved, for example a convenient combined test of a model against current LHC and future ILC results. As soon as these issues are solved, we plan to publish the ILC module as an official \Checkmate{} version on \url{http://checkmate.hepforge.org} and an updated manual.

\section*{Acknowledgements}
This work has been funded by the Collaborative Research Center SFB 676 “Particles, Strings
and the Early Universe” of the Deutsche Forschungsgemeinschaft.

\appendix

\section{Setting up \Checkmate{}}
\label{sec:setup}
\begin{figure}
\begin{Verbatim}[commandchars=\\\@\@,frame=lines,fontsize=\scriptsize]
[Parameters]
Name: VectorDM
Analyses: ILD
Collider: ILC
Sqrts:         500 GeV        
Luminosity:    500 fb-1 | 200 fb-1 | 200 fb-1 | 50 fb-1 | 50 fb-1 
Polarisation:  0:0      | 0.8:-0.3 | -0.8:0.3 | 0.8:0.3 | -0.8:-0.3
InvisiblePIDs: 9000006, -9000006
SLHAFile: [...]/spectrum.slha

[DM]
WhizardSinFile: [...]/whizard.sin
MaxEvents: 10000
\end{Verbatim}
\caption{Example input file for the \Checkmate{} runs used for this study.}
\label{fig:example2}
\end{figure} 

\begin{figure}
\begin{Verbatim}[commandchars=\\\@\@,frame=lines,fontsize=\scriptsize]
model = EffDMVector_UFO (ufo ("[...]/whizard-2.6.0/build/models"))
mVmed = 1900.0
mChi = 200.0
process chichigamma_noisr = "e-", "e+" => "Chi", "Chi~", "a"
compile
cuts = any Pt >= 1.5 GeV [a] and any abs (cos(Theta)) <= 0.998 [a]
[...]
\end{Verbatim}
\caption{\texttt{Whizard} \texttt{.sin} file used for the $e^+ e^- \rightarrow \bar chi \chi \gamma$ Monte Carlo event generation in \Checkmate{}, see Fig.~\ref{fig:example2}.}
\label{fig:example3}
\end{figure} 

In Sec.~\ref{sec:chaplhc} we showed a working input file for \texttt{Checkmate} testing a model at the LHC. As explained in the main text, we needed to extend the  set of accessible parameters to account for new aspects which can be tested at a liner collider. In Fig.~\ref{fig:example2} we show an example for the updated parameter card which has been used for our model tests. Besides some new parameters explained below, we have used a new syntax to allow simultaneous tests of different \emph{collider scenarios}, i.e.\ to test various combinations of luminosity and polarisation and let \Checkmate{} determine the overall strongest bound which can be determined. For our example, we test the following scenarios:

Besides the standard parameters ``\texttt{Name, SLHAFile} and \texttt{MaxEvents} which we already encountered, we have introduced new options explained below:

\begin{description}
\item[\texttt{Collider}] Up to now, \Checkmate{} was only capable of testing models at the LHC. Now, however, it is capable of testing both at the LHC and the ILC. This parameter specifies which collider is supposed to be tested. 
\item[\texttt{Sqrts}] In staged scenarios the ILC may run at different centre-of-mass energies. As explained earlier, analyses may be implemented for different centre-of-mass energies and could be individually tested by changing this parameter. Our example monophoton analysis has only been implemented for $\sqrt{s} = \unit[500]{GeV}$ which is wh only this value is allowed for the \texttt{sqrts} parameter at this stage.
\item[\texttt{Luminosity}] As we want to test different collider scenarios which run different phases with individual polarisation settings, we need to define a set of luminosty values. Here, we make use of the newly implemented \texttt{|} syntax: all settings within one column separated by $\texttt{|}$ corresponds to one \emph{collider scenario}. As stated above, we want to test five such scenarios and therefore we need five values here
\item[\texttt{Polarisation}] As stated above, we want to test the various collider scenarios which use different initial state polarisations. These are stated here, again separated with \texttt{|}. 
\item[\texttt{InvisiblePIDs}] Our model contains a dark matter candidate and its antiparticle which within the Monte Carlo sample will be labelled with the Monte Carlo IDs $\pm$ \texttt{9000006} (these numbers have been predefined by \texttt{FeynRules}). We need to explicitly specify that these IDs correspond to experimentally invisible particles, as otherwise the fast detector simulaton \texttt{Delphes} treats each unknown particle as hadronically interacting per default.
\item[\texttt{WhizardSinFile}] \texttt{Whizard} itself uses so-called \emph{sindarin} input files for its setup. For now, we simply take these files as an input parameter from the user and \texttt{CheckMATE} passes these directly to \texttt{Whizard}. We show the relevant details of this file in Fig.~\ref{fig:example3}: analogously to our \texttt{MG5\_aMC@NLO} example in Fig.~\ref{fig:example}, we specify model and process to be simulated. In our particular scenario, we need to employ cuts to forbid soft or collinear photons which lead to a divergent cross section. Note that these cuts are chosen to be softer than those applied in the event selection, see Sec.~\ref{subsec:analys}, and thus will not affect the signal prediction. 
\end{description}

\bibliographystyle{h-physrev5}
\bibliography{contact}

\begin{thebibliography}{10}

\bibitem{Aad:2014vma}
ATLAS, G.~Aad {\em et~al.},
\newblock JHEP {\bf 05}, 071 (2014), arXiv:1403.5294.

\bibitem{Belanger:2015kga}
G.~Belanger {\em et~al.},
\newblock Phys. Rev. {\bf D91}, 115011 (2015), arXiv:1503.07367.

\bibitem{Alwall:2014hca}
J.~Alwall {\em et~al.},
\newblock JHEP {\bf 07}, 079 (2014), arXiv:1405.0301.

\bibitem{Skands:2003cj}
P.~Z. Skands {\em et~al.},
\newblock JHEP {\bf 07}, 036 (2004), arXiv:hep-ph/0311123.

\bibitem{Degrande:2011ua}
C.~Degrande {\em et~al.},
\newblock Comput. Phys. Commun. {\bf 183}, 1201 (2012), arXiv:1108.2040.

\bibitem{Christensen:2008py}
N.~D. Christensen and C.~Duhr,
\newblock Comput. Phys. Commun. {\bf 180}, 1614 (2009), arXiv:0806.4194.

\bibitem{Alloul:2013bka}
A.~Alloul, N.~D. Christensen, C.~Degrande, C.~Duhr, and B.~Fuks,
\newblock Comput. Phys. Commun. {\bf 185}, 2250 (2014), arXiv:1310.1921.

\bibitem{Staub:2013tta}
F.~Staub,
\newblock Comput. Phys. Commun. {\bf 185}, 1773 (2014), arXiv:1309.7223.

\bibitem{Sjostrand:2007gs}
T.~Sj{\"o}strand, S.~Mrenna, and P.~Z. Skands,
\newblock Comput. Phys. Commun. {\bf 178}, 852 (2008), arXiv:0710.3820.

\bibitem{deFavereau:2013fsa}
DELPHES 3, J.~de~Favereau {\em et~al.},
\newblock JHEP {\bf 02}, 057 (2014), arXiv:1307.6346.

\bibitem{Drees:2013wra}
M.~Drees, H.~Dreiner, D.~Schmeier, J.~Tattersall, and J.~S. Kim,
\newblock Comput. Phys. Commun. {\bf 187}, 227 (2015), arXiv:1312.2591.

\bibitem{Kim:2015wza}
J.~S. Kim, D.~Schmeier, J.~Tattersall, and K.~Rolbiecki,
\newblock Comput. Phys. Commun. {\bf 196}, 535 (2015), arXiv:1503.01123.

\bibitem{Dercks:2016npn}
D.~Dercks {\em et~al.},
\newblock (2016), arXiv:1611.09856.

\bibitem{Brau:2016cym}
J.~Brau {\em et~al.},
\newblock PoS {\bf ICHEP2016}, 062 (2016).

\bibitem{Kilian:2007gr}
W.~Kilian, T.~Ohl, and J.~Reuter,
\newblock Eur. Phys. J. {\bf C71}, 1742 (2011), arXiv:0708.4233.

\bibitem{Moretti:2001zz}
M.~Moretti, T.~Ohl, and J.~Reuter,
\newblock p. 1981 (2001), arXiv:hep-ph/0102195.

\bibitem{Nejad:2014sqa}
B.~Chokoufe~Nejad, T.~Ohl, and J.~Reuter,
\newblock Comput. Phys. Commun. {\bf 196}, 58 (2015), arXiv:1411.3834.

\bibitem{Kilian:2011ka}
W.~Kilian, J.~Reuter, S.~Schmidt, and D.~Wiesler,
\newblock JHEP {\bf 04}, 013 (2012), arXiv:1112.1039.

\bibitem{ALLISON2016186}
J.~Allison {\em et~al.},
\newblock Nuclear Instruments and Methods in Physics Research Section A:
  Accelerators, Spectrometers, Detectors and Associated Equipment {\bf 835},
  186  (2016).

\bibitem{Behnke:2013lya}
H.~Abramowicz {\em et~al.},
\newblock (2013), arXiv:1306.6329.

\bibitem{Habermehl:2017dxh}
M.~Habermehl, K.~Fujii, J.~List, S.~Matsumoto, and T.~Tanabe,
\newblock PoS {\bf ICHEP2016}, 155 (2016), arXiv:1702.05377.

\bibitem{Bartels:2012ex}
C.~Bartels, M.~Berggren, and J.~List,
\newblock Eur. Phys. J. {\bf C72}, 2213 (2012), arXiv:1206.6639.

\bibitem{Dreiner:2012xm}
H.~Dreiner, M.~Huck, M.~Krämer, D.~Schmeier, and J.~Tattersall,
\newblock Phys. Rev. {\bf D87}, 075015 (2013), arXiv:1211.2254.

\bibitem{Aaboud:2017phn}
ATLAS, M.~Aaboud {\em et~al.},
\newblock (2017), arXiv:1711.03301.

\bibitem{Busoni:2013lha}
G.~Busoni, A.~De~Simone, E.~Morgante, and A.~Riotto,
\newblock Phys. Lett. {\bf B728}, 412 (2014), arXiv:1307.2253.

\end{thebibliography}

\end{document}